\documentclass[fleqn,usenatbib]{mnras}

\usepackage{newtxtext,newtxmath}

\usepackage[T1]{fontenc}
\usepackage{ae,aecompl}

\usepackage{graphicx}	
\usepackage{amsmath}	
\usepackage{amssymb}	
\usepackage{booktabs}

\newcommand{\be}{\begin{equation}}
\newcommand{\e}{\end{equation}}
\newcommand{\bear}{\begin{eqnarray}}
\newcommand{\ear}{\end{eqnarray}}

\newcommand{\f}{\frac}
\newcommand{\de}{{\rm d}}

\defcitealias{2015ApJ...813L...8M}{MH15}
\defcitealias{2007ApJ...654..731H}{H07}
\defcitealias{2015A&A...578A..83G}{G15}

\def\nat{Nature}

\def\prd{Phys. Rev. D}
\def\mnras{MNRAS}
\def\apj{ApJ}
\def\apjl{ApJL}
\def\apjs{ApJS}
\def\aap{A\&A}

\def\araa{ARA\&A}
\def\aj{AJ}

\def\physrep{Phys. Rep.}
\def\apss{Ap\&SS}               
\def\pasp{Publ. Astr. Soc. Pac.}

\def\pasj{Pub. Astron. Soc. Japan}

\def\jcap{JCAP}

\title[Reionization in dynamical dark energy models]
{First study of reionization in tilted flat and untilted non-flat dynamical dark energy inflation models}

\author[Mitra, Park, Choudhury \& Ratra]
{Sourav Mitra$^1$\thanks{E-mail: hisourav@gmail.com},~
Chan-Gyung Park$^{2,3}$,~
Tirthankar Roy Choudhury$^4$,~
Bharat Ratra$^3$~\\
$^1$Surendranath College, Department of Physics, 24/2 M. G. Road, Kolkata 700009, India\\
$^2$Division of Science Education and Institute of Fusion Science, Chonbuk National University, Jeonju 54896, South Korea\\
$^3$Department of Physics, Kansas State University, 116 Cardwell Hall, Manhattan, KS 66506, USA\\
$^4$National Centre for Radio Astrophysics, TIFR, Post Bag 3, Ganeshkhind, Pune 411007, India\\
}

\begin{document}
\label{firstpage}
\pagerange{\pageref{firstpage}--\pageref{lastpage}}
\maketitle

\begin{abstract}
We examine the effects of dark energy dynamics and spatial curvature on cosmic reionization
by studying reionization in tilted spatially-flat and untilted non-flat XCDM and $\phi$CDM
dynamical dark energy inflation models that best fit the Planck 2015 cosmic microwave
background (CMB) anisotropy and a large compilation of non-CMB data. We carry out a detailed
statistical study, based on a principal component analysis and a Markov chain Monte Carlo
analysis of a compilation of lower-redshift reionization data, to estimate the uncertainties
in the cosmological model reionization histories. We find that, irrespective of the nature
of dark energy, there are significant differences between the reionization histories of the
spatially-flat and non-flat models. Although both the flat and non-flat models can accurately
match the low-redshift ($z\lesssim6$) reionization observations, there is a clear discrepancy
between high-redshift ($z>7$) Lyman-$\alpha$ emitter data and the predictions from non-flat
models. This is solely due to the fact that the non-flat models have a significantly larger
electron scattering optical depth, $\tau_{\rm el}$, compared to the flat models, which requires
an extended and much earlier reionization scenario supported by more high-redshift ionizing
sources in the non-flat models. Non-flat models also require strong redshift evolution in the photon
escape fraction, that can become unrealistically high ($\gtrsim1$) at some redshifts.
However, $\tau_{\rm el}$ is about 0.9-$\sigma$ lower in the tilted flat $\Lambda$CDM model
when the new Planck 2018 data are used and this reduction will partially alleviate the tension
between the non-flat model predictions and the data.  
\end{abstract}

\begin{keywords}
galaxies: high-redshift -- intergalactic medium -- cosmology: dark ages, reionization, first stars
-- large-scale structure of Universe -- dark energy -- inflation.
\end{keywords}



\section{Introduction}
Assuming that general relativity governs cosmological evolution, a number of different
measurements indicate that about 70\% of the current cosmological energy budget comes
from dark energy, a hypothetical substance responsible for the  observed current accelerated
cosmological expansion \citep[e.g.][and references therein]{2017MNRAS.470.2617A,
2017ApJ...835...26F,2018ApJ...859..101S,2018arXiv180706209P}.
The cosmological constant $\Lambda$ is the simplest dark energy candidate, at least from
a general relativistic perspective, and the cosmological model based on it is known as
$\Lambda$CDM \citep{1984ApJ...284..439P}. This now-standard model assumes flat spatial
geometry, with cold dark matter (CDM) being the second-largest
($\sim26\%$; \citealt{2018arXiv180706209P}) contributor to the current energy budget.
The standard $\Lambda$CDM model is consistent with many current observational constraints
\citep[for reviews of $\Lambda$ and $\Lambda$CDM see][and references therein]
{2008PASP..120..235R,2012CRPhy..13..566M,2018FoPh...48.1446L}. However, current data cannot
rule out slightly curved spatial hypersurfaces or mild dark energy dynamics. In this paper
we examine the effects on cosmic reionization of dark energy dynamics and spatial curvature.    

We use reionization observations to constrain the XCDM ideal fluid dynamical
dark energy parametrization, as well as the physically complete $\phi$CDM dynamical
dark energy model in which dark energy is a scalar field
\citep{1988ApJ...325L..17P,1988PhRvD..37.3406R}.\footnote{For discussions of the
$\phi$CDM model see \citet{2007arXiv0706.1963S}, \citet{2009PhRvD..79j3004Y},
\citet{2010ApJ...714.1347S}, \citet{2013ApJ...766L...7F}, \citet{2013PhLB..726...72F},
\citet{2015arXiv151109317A}, \citet{2017MPLA...3250054S},
\citet{2017ApJ...850..183Z},
\citet{2018arXiv180409350S}, \citet{2018arXiv181008586Y}, \citet{2018arXiv181107513S},
and \citet{2019PhRvD..99d3503T}.} There are a number of recent suggestions that
spatially-flat dynamical dark energy models better fit current observational data than
does the standard spatially-flat $\Lambda$CDM model
\citep[see, e.g.][]{2017RAA....17...50Z,2018arXiv180205571O,2018arXiv180305522P,
2018ApJ...869L...8W,2018ApJ...868...83P,2018arXiv181103505S,2019ChPhC..43b5102Z}.
We also use reionization observations to constrain spatial curvature. There also are
a number of recent suggestions that current data are consistent with very mildly
closed dark energy models \citep{2018ApJ...864...80O,2018ApJ...869...34O,
2018ApJ...866...68O,2018arXiv180100213P,2018arXiv180305522P,2018ApJ...868...83P,
2018arXiv180903598P}.\footnote{For discussions of non-flat cosmological models and
observational constraints on spatial curvature, see \citet{2018MNRAS.477L.122W},
\citet{2018ApJ...856....3Y}, \citet{2019MNRAS.483.1104Q}, \citet{2018MNRAS.480..759R},
\citet{2018ApJ...868...29W}, \citet{2018arXiv181002499D}, \citet{2018arXiv181209100X},
\citet{2019JCAP...01..005M}, \citet{2019PhRvD..99d3522A}, \citet{2019PhRvD..99f3502S},
\citet{2019arXiv190105705Z}, and \citet{2019arXiv190203196R}.} 

Recently \cite{2018ApJ...864...80O,2018ApJ...866...68O,2018ApJ...869...34O,2018arXiv180205571O}
and \cite{2018arXiv180100213P,2018arXiv180305522P,2018ApJ...868...83P,2018arXiv180903598P}
have studied both tilted spatially-flat and untilted 
non-flat $\Lambda$CDM, XCDM and $\phi$CDM inflation models (with 
physically-motivated power spectra for energy density spatial inhomogeneities) 
by using Planck 2015 CMB anisotropy and other non-CMB data. They discovered 
that the non-flat models predict a larger value of the reionization optical 
depth parameter, $\tau_{\rm el}$, which may trigger a serious complication in 
another important aspect of observational cosmology: the epoch of reionization
\citep{LoebBarkana01,BarkanaLoeb01, 2006ARA&A..44..415F,tirth06a,tirth09,2013ASSL..396...45Z,
2014PTEP.2014fB112N,2014arXiv1409.4946F,2016ASSL..423...23L}.
Signatures of reionization are believed to be imprinted in the cosmic 
microwave background radiation, especially through Thomson scattering of 
CMB photons with free electrons, which can be quantified by measuring the 
value of $\tau_{\rm el}$. Assuming a spatially-flat $\Lambda$CDM model, 
\cite{2018arXiv180706209P} 
recently estimated $\tau_{\rm el}$ to be $0.054$, which corresponds to 
instantaneous reionization happening at a mean redshift of $\approx7.7$. 
A lower optical depth is consistent with most observations of high-redshift 
quasars and also explains the observed rapid decrease in Ly$\alpha$ emitters 
(LAEs) number densities at $z\sim7$ \citep{2015MNRAS.446..566M,2015MNRAS.452..261C}.

In our earlier work (\citealt{mitra6}; hereafter Paper I), we explicitly 
showed that the reionization scenario at early epochs is significantly 
different in the tilted flat and untilted non-flat $\Lambda$CDM inflation 
models constrained by Planck 2015 CMB data in combination with BAO 
measurements \citep{2018ApJ...864...80O}. The larger value of $\tau_{\rm el}$ 
for the non-flat
case can cause tension with recent estimates of distant Ly$\alpha$ emitters. 
$\tau_{\rm el}$ for the untilted non-flat XCDM and $\phi$CDM inflation models 
have also been reported to be quite large $\sim0.11-0.12$
\citep{2018arXiv180305522P,2018ApJ...868...83P} and hence these models also 
need to be investigated in light of observations related to cosmic 
reionization. 
In this paper we extend our previous work by now considering dynamical dark 
energy (both the tilted flat and the untilted non-flat XCDM and $\phi$CDM 
inflation models) in data-constrained reionization models. Constraints on the
cosmological parameters and reionization optical depths for these dynamical
dark energy models are taken from the analyses of \cite{2018arXiv180305522P,2018ApJ...868...83P}.
As far as we are aware, this paper presents the first 
detailed statistical analysis on reionization in time-varying dark energy 
models. We also update our previous results for the tilted flat and untilted 
non-flat $\Lambda$CDM inflation models by now using updated constraints 
obtained from a much larger compilation of non-CMB data by \cite{2018arXiv180100213P,2018arXiv180305522P}.

Our paper is organized as follows. In Sec.~\ref{sec:method} we summarize the 
cosmological dynamics and the modeling of cosmic reionization in different 
dark energy scenarios. We also discuss the statistical techniques and cosmic reionization
data used in this work. We present our results in Sec.~\ref{sec:results} 
and summarize the main findings of this paper in Sec.~\ref{sec:conclusions}.

\section{Cosmological models, analysis method, and datasets}
\label{sec:method}

\subsection{Cosmological models}
\label{subsec:demodel}

We study three different pairs of dark energy inflation models, with the dark 
energy modelled as a cosmological constant $\Lambda$ (the $\Lambda$CDM models),
or parametrized by an ideal $\rm X$-fluid with time-varying energy density 
$\rho_{\rm X}$ (the XCDM parametrization), or modelled as a dynamical scalar 
field $\phi$ (the 
$\phi$CDM model). For each dark energy case we separately consider the 
spatially-flat cosmological model and the non-flat (closed) cosmological model 
that best fits the cosmological data we compare these six models to.

For the $\Lambda$CDM model, the Friedmann equation for the Hubble parameter 
as a function of redshift is\footnote{We do not display the photon and neutrino
terms in this and the other Friedmann equations that follow, but their effects 
are accounted for in our computations. In particular we assume three neutrino 
species with one being massive with mass $m_\nu = 0.06$ eV.} 
\begin{equation}
  H(z) = H_0 \sqrt{\Omega_{\rm m}(1+z)^3+\Omega_{\rm k}(1+z)^2+\Omega_\Lambda}
\end{equation}
where $H_0$ is the Hubble constant, $\Omega_{\rm m} = \Omega_{\rm c} + 
\Omega_{\rm b}$ is the present value of the non-relativistic matter
density parameter 
(where $\Omega_{\rm c}$ and $\Omega_{\rm b}$ are the present values of the cold 
dark and baryonic matter density parameters), $\Omega_{\rm k}$ is the current 
value of the spatial curvature density parameter, and $\Omega_\Lambda$ is the 
cosmological constant density parameter. The first model we consider is the 
standard spatially-flat $\Lambda$CDM model \citep{1984ApJ...284..439P} where
$\Omega_{\rm k} = 0$ and $\Omega_\Lambda = 1-\Omega_{\rm m}$. In the non-flat 
$\Lambda$CDM model $\Omega_{\rm k} = 1-\Omega_{\rm m}-\Omega_\Lambda \neq 0$.

As yet there is no totally convincing observational evidence for the dark 
energy density being time independent, so here we also consider two dynamical 
dark energy parameterizations as alternatives to the constant dark energy 
density of the $\Lambda$CDM model. The XCDM model is a widely-used, but 
incomplete, parametrization of dynamical dark energy. Here
dark energy is modelled as an ideal fluid with energy density and pressure 
related through the equation of state $\rho_{\rm X} = w_{\rm X} p_{\rm X}$ and 
the equation of state parameter $w_{\rm X}$ is negative with $w_{\rm X} < -1/3$ 
needed for accelerated cosmological expansion. In this case the Friedmann 
equation is   
\begin{equation}
 H(z)= H_0 \sqrt{\Omega_{\rm m}(1+z)^3 + \Omega_{\rm k}(1+z)^2 + \Omega_{\rm X}(1+z)^{3(1+w_X)}} 
\end{equation}
where $\Omega_{\rm X} = 1-\Omega_{\rm m}-\Omega_{\rm k}$ is the present value of 
the $\rm X$-fluid dark energy density parameter and we consider the flat 
case with $\Omega_{\rm k} = 0$ as well as the closed XCDM model with 
$\Omega_{\rm k} \neq 0$. When $w_{\rm X} = -1$ the 
XCDM parameterization reduces to the physically-complete $\Lambda$CDM model
with $\Omega_{\rm X}=\Omega_\Lambda$.

Although the XCDM parametrization is a widely-used dynamical dark energy 
parameterization, it does not provide a consistent picture for the evolution 
of energy density spatial inhomogeneities.\footnote{Here, when computing the evolution
of spatial inhomogeneities in the XCDM parameterization, we arbitrarily assume that
acoustic disturbances propagate at the speed of light.} The simplest physically complete
dynamical dark energy model is the $\phi$CDM model
\citep{1988ApJ...325L..17P,1988PhRvD..37.3406R,2013PhRvD..88l3513P} which is based on the 
evolution of a rolling scalar field $\phi$ with an inverse-power-law 
potential energy density
\begin{equation}
 V(\phi) = \frac{1}{2}\kappa m_p^2 \phi^{-\alpha}
\end{equation}
where $m_p$ is the Planck mass and $\alpha$ is a positive constant that 
determines the value of the coefficient $\kappa$ 
(see \citealt{1988ApJ...325L..17P,2013PhRvD..88l3513P,2015Ap&SS.357...11F}). 
In the $\phi$CDM model, the Hubble parameter evolves as 
\begin{equation}
 H(z) = H_0 \sqrt{\Omega_{\rm m}(1+z)^3+\Omega_{\rm k}(1+z)^2+\Omega_\phi(z, \alpha)}
\end{equation}
where the time-dependent scalar field dark energy density parameter
\begin{equation}
 \Omega_\phi(z, \alpha) = \frac{1}{6{H_0}^2}\left[\dot{\phi}^2+\kappa m_p^2 \phi^{-\alpha}\right],
\end{equation}
where the overdot denotes the time derivative. $\Omega_\phi(z, \alpha)$ is 
computed from a 
numerical solution of the coupled nonlinear scalar field and Friedmann 
equations of motion. We consider both the closed $\phi$CDM model with 
$\Omega_{\rm k} \neq 0$ as well as the 
spatially-flat case with $\Omega_{\rm k}=0$. In the $\alpha=0$ limit the 
$\phi$CDM model reduces to the $\Lambda$CDM model. 

The primordial power spectra of energy density spatial inhomogeneities in 
these models are determined by quantum fluctuations during an early epoch 
of inflation. The spatially-flat models assume an early epoch of tilted 
non-slow-roll spatially-flat inflation \citep{1985PhRvD..32.1316L,1992PhRvD..45.1913R,1989PhRvD..40.3939R}
with primordial power spectrum
\begin{equation}
   P(k)=A_s \left(\frac{k}{k_0} \right)^{n_s},
\label{eq:Pk}
\end{equation}
where $k$ is wavenumber, the pivot wavenumber $k_0=0.05~\textrm{Mpc}^{-1}$,
and $A_s$ and $n_s$ are the amplitude and spectral 
index. The primordial power spectrum in the untilted slow-roll non-flat 
inflation model \citep{1982Natur.295..304G,1984NuPhB.239..257H,1985PhRvD..31.1931R} is \citep{1995PhRvD..52.1837R,2017PhRvD..96j3534R}
\begin{equation}
   P(q) \propto \frac{(q^2-4K)^2}{q(q^2-K)},
\label{eq:Pq}
\end{equation}
where $q$ is the non-flat space wavenumber and spatial curvature 
$K=-H_0^2 \Omega_{\rm k}$. In the closed, negative $\Omega_{\rm k}$, case, normal
 modes are labeled by $q K^{-1/2}=3,4,5,\cdots$, and the eigenvalue of the 
spatial Laplacian $\propto -(q^2 - K)/K \equiv - {\bar k}^2/K$. $P(q)$ is normalized to $A_s$ at the 
$k_0$ pivot wavenumber. In the $K=0$ spatially-flat limit $P(q)$
reduces to the $n_s=1$ untilted spectrum.

As an aside, we note that the Planck non-flat model analyses \citep{2016A&A...594A..13P,2018arXiv180706209P}
are not based on either of the above power spectra, instead they use
\begin{equation}
   P_{\rm Planck}(q) \propto \frac{(q^2-4K)^2}{q(q^2-K)} \left(\frac{\bar k}{k_0} \right)^{n_s-1}, 
\label{eq:PqP}
\end{equation}
where in addition to the non-flat space wavenumber $q$, the wavenumber 
$\bar k$ is also used to define and tilt the non-flat model 
$P(q)$. The ${\bar k}^{n_s-1}$ tilt factor in $P_{\rm Planck}(q)$ assumes that 
tilt in non-flat space works somewhat as it does in flat space, which seems 
unlikely since spatial curvature sets an additional length scale in 
non-flat space (i.e., in addition to the Hubble length). It is not known if the 
power spectrum of Eq.~(\ref{eq:PqP}) can be the consequence of quantum 
fluctuations during an early epoch of inflation. This power spectrum is 
physically sensible if $K= 0$ or if $n_s = 1$, when it reduces to the 
power spectra in Eqs.~(\ref{eq:Pk}) and (\ref{eq:Pq}), both of which are 
consequences of quantum fluctuations during inflation.

Constraints on cosmological parameters can be obtained by performing a 
Monte-Carlo Markov chain (MCMC) analysis over the corresponding cosmological 
model parameter space for a combination of CMB and non-CMB data.
Building on the work of \cite{2018ApJ...864...80O}, \citet{2018arXiv180100213P}
have analyzed the six-parameter tilted flat and untilted non-flat $\Lambda$CDM 
inflation models with the power spectra of Eqs.~(\ref{eq:Pk}) and 
(\ref{eq:Pq}). The tilted flat model is conventionally parameterized 
by $\Omega_{\rm b}h^2, \Omega_{\rm c}h^2, \theta, \tau_{\rm el}, A_s$ and $n_s$
while the untilted non-flat model uses $\Omega_{\rm k}$ instead of $n_s$. Here
$h$ is the Hubble constant in units of 100 km s$^{-1}$ Mpc$^{-1}$ and $\theta$ is
the angular diameter distance as a multiple of the acoustic Hubble radius at 
recombination. For these analyses, \citet{2018arXiv180100213P} used Planck 
2015 CMB anisotropy data \citep{2016A&A...594A..13P} and a number of non-CMB 
datasets. Similar analyses have been performed for the seven parameter 
XCDM \citep{2018arXiv180305522P} and $\phi$CDM \citep{2018ApJ...868...83P} 
dynamical dark energy inflation models, with $w_{\rm X}$ and $\alpha$, 
respectively, being the seventh parameter.
In this paper we used their results to constrain reionization scenarios in 
the six cosmological models, tilted spatially-flat or untilted non-flat, and 
with constant or dynamical dark energy density. 

We note that unlike the Planck 2015 and 2018 analyses of a seven parameter 
tilted non-flat 
$\Lambda$CDM model with the power spectrum of Eq.~(\ref{eq:PqP}) that favors 
flat geometry \citep{2016A&A...594A..13P,2018arXiv180706209P}, an analysis 
of the six parameter untilted non-flat $\Lambda$CDM inflation model with the 
power spectrum 
of Eq.~(\ref{eq:Pq}) favors a very mildly closed model at more than 5-$\sigma$
\citep{2018arXiv180100213P}.

In the spatially-flat case, \citet{2018arXiv180205571O} found that the best-fit
seven parameter tilted flat XCDM and $\phi$CDM inflation models had a slightly lower 
$\chi^2$ than the best-fit six parameter tilted flat $\Lambda$CDM model. This 
was 
confirmed by \citet{2018arXiv180305522P}, \citet{2018ApJ...868...83P}, and \citet{2018arXiv181103505S}.
However, in both best-fit models, dark energy was not 
inconsistent with a cosmological constant. In all three best-fit untilted 
non-flat cases, $\chi^2$ is an additive factor of 10---20 larger (depending on data combination 
used) than in the  best-fit six parameter tilted flat $\Lambda$CDM model. 
However, the six parameter tilted flat $\Lambda$CDM model does not nest 
inside any of the three untilted non-flat models and so it is not possible to turn these $\chi^2$ 
differences into goodness-of-fit probabilities.

In Table \ref{tab:params} we have listed the best-fit mean values of 
cosmological parameters for the flat and non-flat $\Lambda$CDM, XCDM and 
$\phi$CDM models (i.e.\ six different cases) as obtained from 
MCMC analyses using Planck 2015 TT+ lowP + lensing CMB anisotropy 
\citep{2016A&A...594A..13P} and SNIa, BAO, $H(z)$, and growth rate 
$f(z)\sigma_8(z)$ data. For detailed discussions of the method of analyses and the data 
used, see \citet{2018ApJ...864...80O,2018ApJ...869...34O,2018ApJ...866...68O,2018arXiv180205571O}
and \citet{2018arXiv180100213P,2018arXiv180305522P,2018ApJ...868...83P}.

\begin{table*}
\begin{center}
\def\arraystretch{1.2}%
\makebox[\textwidth][c]{
    \begin{tabular}{l|ccc|ccc}
\toprule
\toprule
Parameter & \multicolumn{3}{c}{Tilted flat models} & \multicolumn{3}{c}{Untilted non-flat models}\\
\cmidrule(lr){2-4}\cmidrule(lr){5-7}
& $\Lambda$CDM & XCDM & $\phi$CDM & $\Lambda$CDM & XCDM & $\phi$CDM\\
\midrule
$\Omega_{\rm b}h^2$ & $0.02232$ & $0.02233$ & $0.02238$ & $0.02305$ & $0.02305$ & $0.02304$\\
$\Omega_{\rm c}h^2$ & $0.1177$ & $0.1175$ & $0.1168$ & $0.1093$ & $0.1092$ & $0.1093$\\
$\Omega_{\rm k}$ & --- & --- & --- & $-0.0083$ & $-0.0069$ & $-0.0063$\\
$h$ & $0.6919$ & $0.6806$ & $0.6763$ & $0.6801$ & $0.6745$ & $0.6736$\\
$\sigma_8$ & $0.8117$ & $0.8103$ & $0.8055$ & $0.8121$ & $0.8055$ & $0.8051$\\
$n_s$ & $0.9692$ & $0.9696$ & $0.9715$ & --- & --- & ---\\
$w_{\rm X}$ & --- & $-0.994$ & --- & --- & $-0.960$ & ---\\
$\alpha$ & --- & --- & $<0.22$ & --- & --- & $<0.31$\\
\midrule
$\tau_{\rm el}$ & $0.066\pm0.012$ & $0.068\pm0.015$ & $0.074\pm0.014$ & $0.112\pm0.012$ & $0.119\pm0.012$ & $0.122\pm0.012$\\
\bottomrule
\bottomrule
\end{tabular}
}
\caption{{\it Upper rows:} Best-fit mean values of the cosmological parameters for tilted flat and untilted non-flat
$\Lambda$CDM (from \citealt{2018arXiv180305522P}), XCDM (from \citealt{2018arXiv180305522P}), and $\phi$CDM
(from \citealt{2018ApJ...868...83P}) inflation models constrained using Planck 2015 TT + lowP + lensing CMB anisotropy and SNIa, BAO, $H(z)$
and growth rate data. The uncertainties in these parameters have not been considered
in our analyses here. {\it Bottom row:} electron scattering optical depths, $\tau_{\rm el}$, for the corresponding model
(mean and $68.3\%$ confidence limits), which we use in the present analysis to constrain reionization parameters.}
\label{tab:params}
\end{center}
\end{table*}

Note that, except for the reionization optical depth $\tau_{\rm el}$, here we use only the mean values for all other
cosmological parameters and neglect their uncertainties. However, we did a thorough check by considering the corresponding
$\pm1$-$\sigma$ errors around the mean value of each parameter at a time, keeping the others fixed at their central values,
and found that ignoring the uncertainties or correlations between the 
cosmological parameters does not make much of a difference in our final 
results. This is because of the fact that the cosmic reionization model 
itself has many assumptions and uncertainties, as we will 
soon see. Perhaps the most significant parameter related to reionization is the electron scattering
optical depth $\tau_{\rm el}$. Its mean values along with $68.3\%$ (1-$\sigma$) confidence limits (C.L.)
for the six different models are quoted in the bottom row of Table \ref{tab:params}. We have used these mean
values and uncertainties in our analysis to constrain reionization parameters.
Since $\tau_{\rm el}$ has the most significant effect, we emphasize that the 
Planck 2018 \citep{2018arXiv180706209P} estimate in the six parameter 
tilted flat $\Lambda$CDM inflation model is $\tau_{\rm el} = 0.054 \pm 0.007$, 
about 0.9-$\sigma$ (of the quadrature sum of the two error bars) lower than 
the corresponding $\tau_{\rm el} = 0.066 \pm 0.012$ (last row of the second 
column of Table \ref{tab:params}) used here.

Also note that, in order to compute the star formation history for dynamical 
dark energy models, one needs to use appropriate
values for the linear growth factor of dark matter perturbations, $D(z)$, and 
the rms mass fluctuation $\sigma(M)$, at mass scale
$M$ \citep{2001MNRAS.322..419H,2003ApJ...599...24M}, the latter being computed 
by integrating the corresponding power spectrum $P(k)$. In this paper both 
of these quantities are computed from the best-fit parameter value results of
\cite{2018arXiv180305522P,2018ApJ...868...83P}.

\subsection{Modeling cosmic reionization}
\label{subsec:cfmodel}

We use a semi-analytical approach to model cosmic reionization in order to 
constrain the various inflation scenarios presented above. The main features 
of this model are based on the work of \cite{tirth05,tirth06}. We refer the 
reader to these papers for a detailed description and in what follows we
summarize the procedure.

The IGM density field is assumed to have a lognormal distribution at low densities and changes to a power-law
form at high densities \citep{tirth05}. The model takes into account the inhomogeneities in the IGM appropriately
by adopting the method outlined in \cite{2000ApJ...530....1M} in which reionization is complete once all
the low-density regions are ionized. The denser regions remain neutral for a 
longer time due to their high recombination rate \citep{tirth09}. The mean 
free path of ionizing photons is computed from the distribution of these 
high density regions as \citep{tirth06}
\be
 \lambda_{\rm mfp}(z) = \f{\lambda_0}{[1 - F_V(z)]^{2/3}}, 
 \label{eq:mfp}
\e
where $F_V$ is the volume fraction of ionized regions and $\lambda_0$ is a normalization constant
which we treat as a free parameter in our model. The parameter $\lambda_0$ is usually constrained
from the low-redshift observations on the number of Lyman limit systems (LLS) per unit redshift range,
$\de N_{\rm LL}/\de z$, which can be computed in our model from the evolution of the mean free path
\be
 \f{\de N_{\rm LL}}{\de z}
 = \f{c}{\sqrt{\pi}~\lambda_{\rm mfp}(z) H(z) (1+z)}.
 \label{eq:LLS}
\e

Although there should be a dependence on how far a
ionizing source is from these regions, we do not take that into account 
in this simplified model. Also, we assume that the photons will be absorbed 
``locally'', right after being emitted, which is a reasonable
approximation for describing hydrogen reionization, particularly 
when $z\gtrsim3$
\citep{1999ApJ...514..648M,2000ApJ...530....1M,tirth09}. Moreover, these 
approximations work quite well when studying global properties of 
reionization and matching these against the current data, which is what is 
considered in this work. The thermal and ionization history of the universe is 
computed self-consistently incorporating radiative feedback (UV photons from 
stars could increase the minimum mass for star-forming haloes in the ionized 
regions and hence could influence the subsequent star formation history;
\citealt{tirth05,2008MNRAS.390..920O,2013MNRAS.432.3340S}) in the model.

In this model, reionization is assumed to be driven by two types of sources:
(i) Pop II stars with a Salpeter IMF in the mass range 1---100 $M_\odot$, and 
(ii) quasars. Although at lower redshifts quasars have been considered as 
significant ionizing sources, they have negligible contribution to the UV 
ionizing background at $z\gtrsim6$ (\citealt{2017MNRAS.468.4691D,mitra5,2018MNRAS.473..227H};
but also see \citealt{2015ApJ...813L...8M,2016MNRAS.457.4051K} for QSO-driven reionization models).
The model incorporates the QSO contribution by computing their ionizing emissivities based on the observed
luminosity functions at $z<6$ \citep{2007ApJ...654..731H}. Note that we do 
not consider here other sources of ionizing photons such as Pop III stars,
exotic particles like decaying dark matter candidates etc.,
as the current constraints on such objects make it improbable that they could reionize the IGM by themselves
\citep{2013ASSL..396...45Z}. As there is only one type of stellar population 
in our model, there is no need to include a direct chemical feedback effect 
(stars expel metals into the medium and change its chemical composition and 
that affects subsequent star formation; \citealt{tirth09}) here. However,
such an effect can be indirectly incorporated in our model by a method 
described in the next section. 

The rate of ionizing photons produced from 
star-forming haloes is computed from \citep{tirth05,tirth09}
\be
 \dot{n}_{\rm ph} = N_{\rm ion} n_{\rm b} \f{\de f_{\rm coll}}{\de t}.
\e
Here $f_{\rm coll}$ is the fraction of mass that has collapsed into halos, 
computed using an appropriate halo mass function \citep{1974ApJ...187..425P}, 
$n_{\rm b}$ is the total baryonic number density, and $N_{\rm ion}$ is the number of 
ionizing photons per baryon produced by Pop II stars, which is often  parametrized as
\citep{tirth09,mitra3,mitra4}
\be
 N_{\rm ion}=\epsilon_*f_{\rm esc}N_\gamma,
 \label{eq:Nion}
\e
where $\epsilon_*$ is the star-forming efficiency, $f_{\rm esc}$ is the 
fraction of UV photons escaping into the IGM, and
$N_\gamma$ is the specific number of photons emitted per baryon in stars.
Once $\dot{n}_{\rm ph}$ is known, we can compute the photoionization rate ($\Gamma_{\rm PI}$)
using the relation
\be
 \Gamma_{\rm PI}(z) = 
 (1 + z)^3 \int_{\nu_{\rm HI}}^{\infty} \de \nu ~ \lambda_{\rm mfp}(z;\nu) 
 \dot{n}_{\rm ph}(z; \nu) \sigma_H(\nu) 
\e
where $\nu_{\rm HI}$ is the threshold frequency for photoionization of hydrogen and 
$\sigma_H(\nu)$ is the photoionization cross section.

\subsection{Datasets, parameters, and the MCMC + PCA method}
\label{subsec:mcmc}

It is very likely that the parameter $N_{\rm ion}$ depends on halo mass ($M$) 
and redshift ($z$) but, due to our limited understanding of complex 
star-formation physics, modeling it as a function of $M$ and $z$
still remains as an unsettled issue (\citealt{2006MNRAS.371L...1I,2018MNRAS.475.3870S}, but also see
\citealt{2019MNRAS.tmp...37P}). However, it is possible to find  $N_{\rm ion}$ as a function of $z$ with help
of a method called principal component analysis or PCA. A brief description of 
this approach follows.

PCA is a robust and widely used technique to analyze data by constructing a new set of eigenvectors
(also known as principal components) which are optimized to describe noisy datasets by using the fewest
number of components but without losing significant information. It has been implemented for the analysis
of several astrophysical and cosmological systems, see \citet{1999MNRAS.304...75E}, \citet{2003PhRvD..68b3001H},
\citet{2003PhRvL..90c1301H}, \citet{2006MNRAS.372..646L}, \citet{2008ApJ...672..737M}, \citet{2010PhRvL.104u1301C},
\citet{2011A&A...527A..49I}, \citet{2012PASP..124.1015B}, \citet{2012MNRAS.421.3570G}, \citet{2015MNRAS.448.2232R},
\citet{2015PhRvD..91f3514M}, and \citet{2018ApJ...863..173M} for discussions 
and references. Following \cite{mitra1,mitra2}, we parametrize $N_{\rm ion}$ as 
an arbitrary function of $z$ by a set of $n_{\rm bin}$ discrete free parameters
with redshift bin width $\Delta z=0.2$, and decompose $N_{\rm ion}(z)$ into its principal components
\begin{equation}
 N_{\rm ion}(z)=N_{\rm ion}^{\rm fid}(z)+\sum_{k=1}^{n_{\rm bin}} m_kS_k(z) .
 \label{eq:PCA}
\end{equation}
Here $S_k(z)$, also known as the principal components, are the eigenfunctions of the Fisher information matrix
that expresses the dependence of the observed datasets on $N_{\rm ion}(z)$, 
and $m_k$ are the expansion coefficients or amplitudes of the principal components.
The Fisher matrix is constructed as \citep{mitra1,mitra2}:
\begin{equation}
 F_{ij}=\sum_{\alpha=1}^{n_{\rm obs}}\frac{1}{\sigma_{\alpha}^2}
\frac{\partial {\cal G}_{\alpha}^{\rm th}}{\partial N_{\rm ion}^{\rm fid}(z_i)}
\frac{\partial {\cal G}_{\alpha}^{\rm th}}{\partial N_{\rm ion}^{\rm fid}(z_j)},
\label{eq:fisher}
\end{equation}
where ${\cal G}_{\alpha}$ ($\alpha=1,2,\ldots,n_{\rm obs}$) represents the set of $n_{\rm obs}$ number of
observational data points (described below) with their corresponding errors $\sigma_{\alpha}$.
${\cal G}_{\alpha}^{\rm th}$ is the theoretical value of ${\cal G}_{\alpha}$.
$N_{\rm ion}^{\rm fid}(z)$ is the fiducial model at which the Fisher
matrix is computed and is chosen  in such a way that it can produce at least a reasonable match with all the observed
data considered here at $z<6$ and also leads to an acceptable $\tau_{\rm el}$ for different models. In our analysis here
we have chosen a constant $N_{\rm ion}^{\rm fid}=15$ for all the flat models, whereas an evolving $N_{\rm ion}^{\rm fid}(z)$
at higher redshifts ($z>6$) is assumed for the non-flat models in order to achieve higher $\tau_{\rm el}$ values for these.
We should emphasize here that, although our true or actual underlying $N_{\rm ion}(z)$ is slightly different
from the fiducial model, the main conclusions of this work will hold true for any choice of $N_{\rm ion}^{\rm fid}(z)$
as long as it reasonably matches all the observations. 

The observational data (${\cal G}_{\alpha}$) used here to construct the Fisher matrix are:
\begin{enumerate}
 \item hydrogen photoionization rates $\Gamma_{\rm PI}$ in the range $2.4\leqslant z\leqslant6$ from \cite{2011MNRAS.412.1926W}
  and \cite{2013MNRAS.436.1023B}. These data points are based on observations of mean opacity of the IGM to Ly$\alpha$ photons
  and the IGM temperature. Note that their computation of ionization rates somewhat depends on the adopted cosmological
  and astrophysical parameters, which we have accounted for in our work here.
 \item distribution of Lyman limit systems (LLS) $\de N_{\rm LL}/\de z$ over a redshift range of $2<z<6$ from the combined
  datasets of \cite{2010ApJ...721.1448S} and \cite{2010ApJ...718..392P}.
 \item reionization optical depths $\tau_{\rm el}$ as obtained from the tilted flat and untilted non-flat $\Lambda$CDM
  \citep{2018arXiv180305522P}, XCDM \citep{2018arXiv180305522P}, and $\phi$CDM \citep{2018ApJ...868...83P} inflation models.
  We have listed these values in the bottom row of Table \ref{tab:params}.
  In our model, this quantity is directly computed from the global average value of the comoving free electron density,
  $n_e(z)$, as $\tau_{\rm el}(z) = \sigma_T c \int \de t ~ n_{e}$ (with $\sigma_T$ being the Thomson scattering cross section).
\end{enumerate}
We emphasize that we keep all other cosmological parameters, corresponding to the different models,
at their best-fit mean values, as listed in the upper rows of Table \ref{tab:params}. Ideally, one can include
their uncertainties in the reionization model and vary those as free parameters, but that would require a more
complicated analysis as well as significantly more computational resources and so is beyond the 
scope of this paper. Thus the uncertainties in reionization history presented
here are slightly smaller than they really are.

One crucial advantage of using the principal component method instead of 
some other parameterization is that most of the significant information is 
in the first few eigenmodes associated with larger eigenvalues and these
are the most accurately measured modes with the smallest uncertainties. 
This means that we can retain only the first few terms, say up to $M$ 
(where $M<n_{\rm bin}$), in the sum over $k$ in Eq.~(\ref{eq:PCA}) and discard 
the other modes which are noisier (having smaller eigenvalues) and contribute 
less to the reconstruction of $N_{\rm ion}(z)$:
\begin{equation}
 N_{\rm ion}(z)=N_{\rm ion}^{\rm fid}(z)+\sum_{k=1}^{M} m_kS_k(z) .
 \label{eq:PCA2}
\end{equation}
We have checked that in our case modes with $M>7$ would produce hopelessly large errors in the recovered 
quantities and thus can be safely discarded. Once we have the optimal number 
of eigenmodes, the next step is to employ a thorough MCMC 
analysis\footnote{To generate the chains of Monte Carlo samples, we have developed a code based
on the publicly available \texttt{CosmoMC} package (\citealt{2002PhRvD..66j3511L}; http://cosmologist.info/cosmomc/).} 
over the parameter space of the corresponding PCA amplitudes ($m_k$) and the mean free path normalization
constant ($\lambda_0$) in order to get constraints on $N_{\rm ion}(z)$ and other quantities related to
reionization history.
To achieve accurate results from the MCMC analyses, we run enough separate parameter chains until
the Gelman and Rubin convergence statistics satisfies $R-1<0.01$, where $R$ is the ratio
of variance of parameters between chains to the variance within each chain.
The likelihood function we are maximizing (corresponding to $\chi^2$-minimization) here is:
\be
 L \propto \exp\left(-\frac{\chi^2}{2}\right) = \exp \left(-\f{1}{2}\sum_{\alpha=1}^{n_{\rm obs}}
\f{\left({\cal G}_{\alpha}^{\rm obs} - {\cal G}_{\alpha}^{\rm th}\right)^2}{\sigma_{\alpha}^2}
\right)
 \label{eq:likelihood}
\e
where ${\cal G}_{\alpha}^{\rm obs}$ are the above-mentioned observational data points
i.e., ${\cal G}_{\alpha} = \{\Gamma_{\rm PI},~ \de N_{\rm LL}/\de z,~ \tau_{\rm el}\}$
and $\sigma_{\alpha}$ are their corresponding errors. Note that, here the $m_k = \{m_1,\ldots,m_M\}$
(along with $\lambda_0$) are the parameters we fit
at the MCMC stage. We can now describe our model by these coefficients instead of the original
parameter $N_{\rm ion}(z)$ as they ($m_k$) have the crucial advantage of being uncorrelated.\footnote{However, there could be a correlation between $m_k$ and $\lambda_0$, but that is
irrelevant for the purpose of the study presented here.}
The $N_{\rm ion}(z)$ and other quantities, e.g., $\Gamma_{\rm PI}(z)$, $\tau_{\rm el}(z)$,
neutral hydrogen fraction $x_{\rm HI}(z)$ etc., can then be obtained from marginalized posterior distributions determined using the fitted parameters.

To avoid the inherent bias which might exist in any particular choice of $M$, we perform repeated MCMC analyses
for all the PCA amplitudes from $M=2$ to $M=7$ and combine their errors together at the final stage (for details, see
\citealt{2010PhRvL.104u1301C,mitra2}). We further impose a model-independent prior on the neutral fraction
$x_{\rm HI}$ of $<0.11$ ($<0.09$) at $z=5.9$ ($5.6$), obtained from \cite{2015MNRAS.447..499M}, as upper limits
at the Monte Carlo stage.

\section{Results and discussion}
\label{sec:results}

\subsection{Constraints on \texorpdfstring{$\Lambda$}{}CDM models}
\label{subsec:results-LCDM}

\begin{figure*}
\centering
  \includegraphics[height=0.5\textwidth, angle=0]{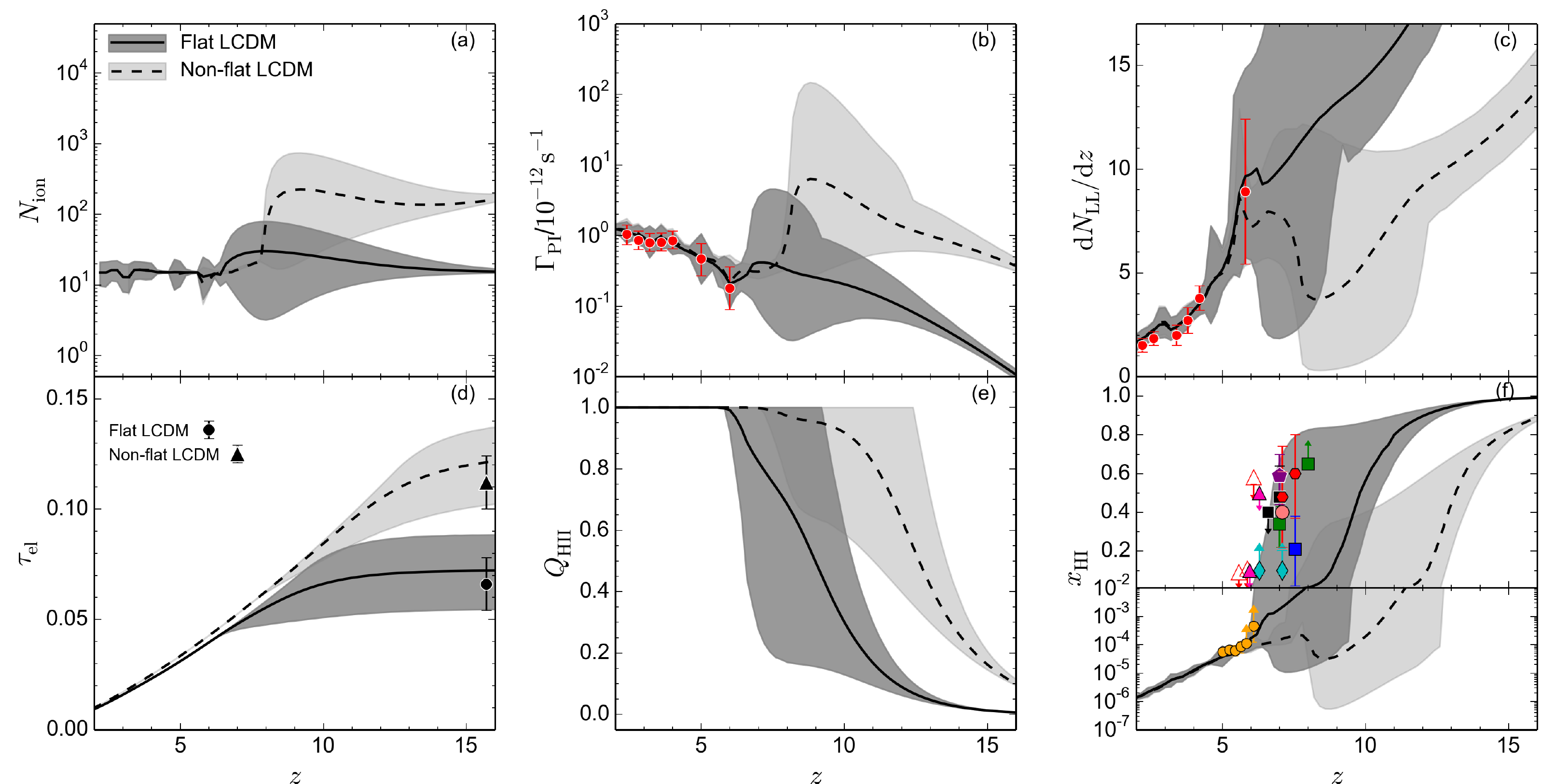}
  \caption{MCMC constraints on various quantities related to reionization obtained from the principal component analysis
  for tilted flat and untilted non-flat $\Lambda$CDM inflation models that best fit the Planck 2015 TT + lowP + lensing and SNIa, BAO, $H(z)$
  and growth rate data \citep{2018arXiv180305522P}. The thick central lines along with surrounding shaded regions correspond to the
  best-fit models and their 2-$\sigma$ uncertainty ranges. The Panels show as a function of redshift:
  (a) number of ionizing photons in the IGM per baryon in stars,
  (b) photoionization rates for hydrogen along with observed data from \citet{2011MNRAS.412.1926W} and \citet{2013MNRAS.436.1023B},
  (c) specific number of Lyman-limit systems with the data points combined from \citet{2010ApJ...721.1448S} and
  \citet{2010ApJ...718..392P},
  (d) electron scattering optical depths along with their values from \citet{2018arXiv180305522P},
  (e) volume filling factor of ionized regions, and
  (f) global neutral hydrogen fraction with various current observational limits. See the main text for references.}
\label{fig:MCMC-LCDM}
\end{figure*}

The MCMC results for the tilted flat and untilted non-flat $\Lambda$CDM 
inflation models are shown in Fig.~\ref{fig:MCMC-LCDM}. The thick central 
solid lines and dark shaded regions surrounding them correspond to the mean 
values and 2-$\sigma$ (95\% C.L.) uncertainty ranges, respectively, of 
different quantities related to reionization history for the standard tilted 
flat $\Lambda$CDM model. On the other hand, the thick central dashed lines 
and light shaded regions represent the same for the untilted non-flat 
$\Lambda$CDM model. The MCMC constraints on all these quantities are much 
tighter for redshift $z\lesssim6$, as most of the reionization-related
datasets considered in our analysis exist only at these redshifts. However, 
evolution in the $z>6$ region essentially depends on the optical depth data ($\tau_{\rm el}$) 
alone, and that's why a weaker constraint is expected in this redshift regime.
The errors further decrease at $z>12$ as the components of the Fisher matrix
are zero at higher redshifts due to the fact that there exist no free electrons to contribute to
$\tau_{\rm el}$ at that regime, and thus forcing the mean model to converge towards the
fiducial one. Also note that the evolution is almost identical for the flat and non-flat 
cases at $z\lesssim6$, however, at earlier epochs their evolutionary histories 
differ considerably, as expected from the significantly different values of 
optical depths in the two models. 

The evolution of $N_{\rm ion}(z)$ obtained from the MCMC statistics is shown in 
Panel (a). We see that an almost constant or non-evolving $N_{\rm ion}$ is sufficient to explain
the reionization history in flat $\Lambda$CDM, whereas this
quantity must increase at $z\gtrsim7$ for the non-flat model. This is due to the fact that, the value of $\tau_{\rm el}$
is quite high ($\sim0.11$) for this case, and $N_{\rm ion}$ has to evolve quite 
rapidly at higher redshifts to match such a value. This hints that either 
chemical feedback from Pop III stars and/or an evolving star-forming efficiency or photon
escape fraction or both play a significant role in the non-flat $\Lambda$CDM model. We have shown the evolution of
$\Gamma_{\rm PI}$ and $\de N_{\rm LL}/\de z$ in Panels (b) and (c) respectively along with their
observed data (red points with error bars) that we have included in our MCMC analysis. Clear non-monotonic
trends with redshift are also visible here for the non-flat case. $\Gamma_{\rm PI}$ is quite large at earlier
epochs in this model compared to that for the flat one, indicating a possibility of Pop III stars being the
major contributers of ionizing photons at those epochs.
As expected, the $\tau_{\rm el}$, shown in Panel (d), reasonably match the current data points from the results of
\cite{2018arXiv180305522P}. However, its upper limits for both the flat and non-flat models are found to be slightly
higher than their observed values, suggesting that a wide range of early reionization models are still permitted
by the data. Panel (e) shows the evolution of volume filling factor of ionized (HII) region, $Q_{\rm HII}(z)$ (defined
as the fraction of IGM volume that is filled up by ionized regions). From 
this plot one can see that for the tilted flat $\Lambda$CDM model the mean 
growth of $Q_{\rm HII}(z)$ is quite smooth and reionization is almost completed,
i.e. $Q_{\rm HII}(z)\sim1$, around $6\lesssim z\lesssim9$ (95\% C.L.). On the 
other hand the untilted non-flat model shows a much faster
rise at initial stages starting as early as $z\approx14$ and then gradually approaches towards the end-stage of reionization
which completes around $7\lesssim z\lesssim12$ (95\% C.L.). A more extended reionization scenario is needed for the
non-flat case so that enough contribution to $\tau_{\rm el}$ is acquired in 
order to match the high value.

A similar conclusion can also be drawn from the evolution of the neutral 
hydrogen fraction $x_{\rm HI}(z)$ in Panel (f).
For comparison, here we have shown various recent observational limits on this quantity. The most important constraints
at the end-stage of cosmic reionization come from the Gunn-Peterson optical depth data of high-redshift ($z\approx5-6$)
quasars measured by \cite{2006AJ....132..117F} (filled yellow circles). Measurements from the near zone of bright quasars
by \cite{2013MNRAS.428.3058S} and \cite{2011MNRAS.416L..70B} are shown in the figure by filled cyan diamonds. Constraints
on the neutral fraction from the damping wing analysis of highest redshift ($z=7.09$ and $7.54$) quasars by \cite{2017MNRAS.466.4239G}
and \cite{2018ApJ...864..142D} are also shown here by the filled salmon circle and red hexagons respectively. Recently,
a more model-independent analysis by \cite{2019MNRAS.484.5094G} on the $z = 7.54$ QSO recovers a slightly lower value of
$x_{\rm HI}=0.21^{+0.17}_{-0.19}$ (1-$\sigma$ error; shown by the filled blue square in the figure) than that reported
in \cite{2018ApJ...864..142D}. We show the constraints from observed GRB host galaxies at $z\sim6.3$ \citep{2006PASJ...58..485T}
and $z\sim5.9$ \citep{2013ApJ...774...26C} by filled pink triangles. Apart from quasars and GRBs, the high-redshift Lyman$\alpha$
emitters (LAEs) are also a reliable probe for the epoch of reionization. In the plot, we show them by filled black squares
\citep{2008ApJ...677...12O,2010ApJ...723..869O}, green squares \citep{2014ApJ...795...20S}, and a filled purple pentagon
\citep{2018ApJ...856....2M}. Note that most of the observational constraints at $z\gtrsim7$ are extremely model-dependent
and might get modified in the future, that's why we did not include those in our analysis.\footnote{However, see
our Paper I \citep{mitra6}, where we did a separate analysis that explicitly included one of the high-$z$ $x_{\rm HI}$
constraints from LAEs.} A more useful constraint for us comes from a model-independent analysis of high redshift
($z\sim5.6$ and $5.9$) quasar spectra by \cite{2015MNRAS.447..499M} (open red triangles) which we impose in our current
MCMC analysis as priors. One can see that the flat model can comfortably accommodate all these observational constraints on
$x_{\rm HI}$, considering its 2-$\sigma$ region, whereas the non-flat model with larger $\tau_{\rm el}$ value struggles
to match these limits. In fact, any reionization model that produces a higher optical depth results in a smaller
neutral fraction at earlier times (see e.g., \citealt{2013ApJ...768...71R,2015ApJ...802L..19R,2015ApJ...811..140B,mitra4,mitra6}).
We note, however, that the Planck 2018 $\tau_{\rm el}$ value in the six parameter tilted flat 
$\Lambda$CDM inflation model is about 0.9-$\sigma$ lower than 
the corresponding $\tau_{\rm el}$ value we use here; accounting for this will 
alleviate some of the discrepancy between the untilted non-flat $\Lambda$CDM 
model predictions and the observations.   

\subsection{Constraints on dynamical dark energy models}
\label{subsec:results-XCDM-QCDM}

\begin{figure*}
\centering
  \includegraphics[height=0.5\textwidth, angle=0]{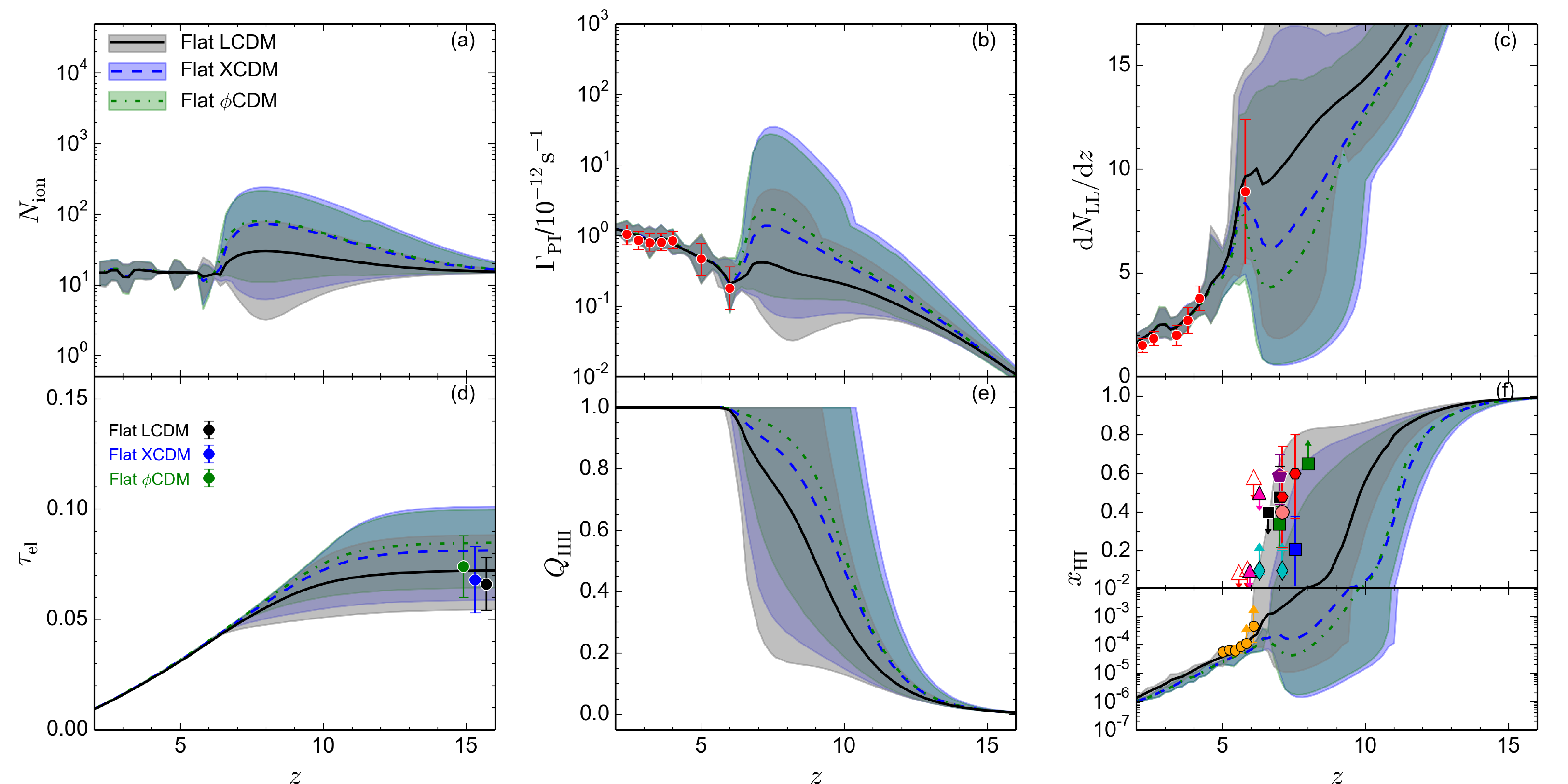}
  \caption{Same as Fig.~\ref{fig:MCMC-LCDM}, but now showing the flat $\Lambda$CDM (solid central lines with
  surrounding gray shaded 2-$\sigma$ regions), flat XCDM (\citealt{2018arXiv180305522P}; dashed central lines
  with surrounding blue shaded 2-$\sigma$ regions), and flat $\phi$CDM (\citealt{2018ApJ...868...83P};
  dot-dashed central lines with surrounding green shaded 2-$\sigma$ region) inflation models.}
\label{fig:MCMC-flat}
\end{figure*}

\begin{figure*}
\centering
  \includegraphics[height=0.5\textwidth, angle=0]{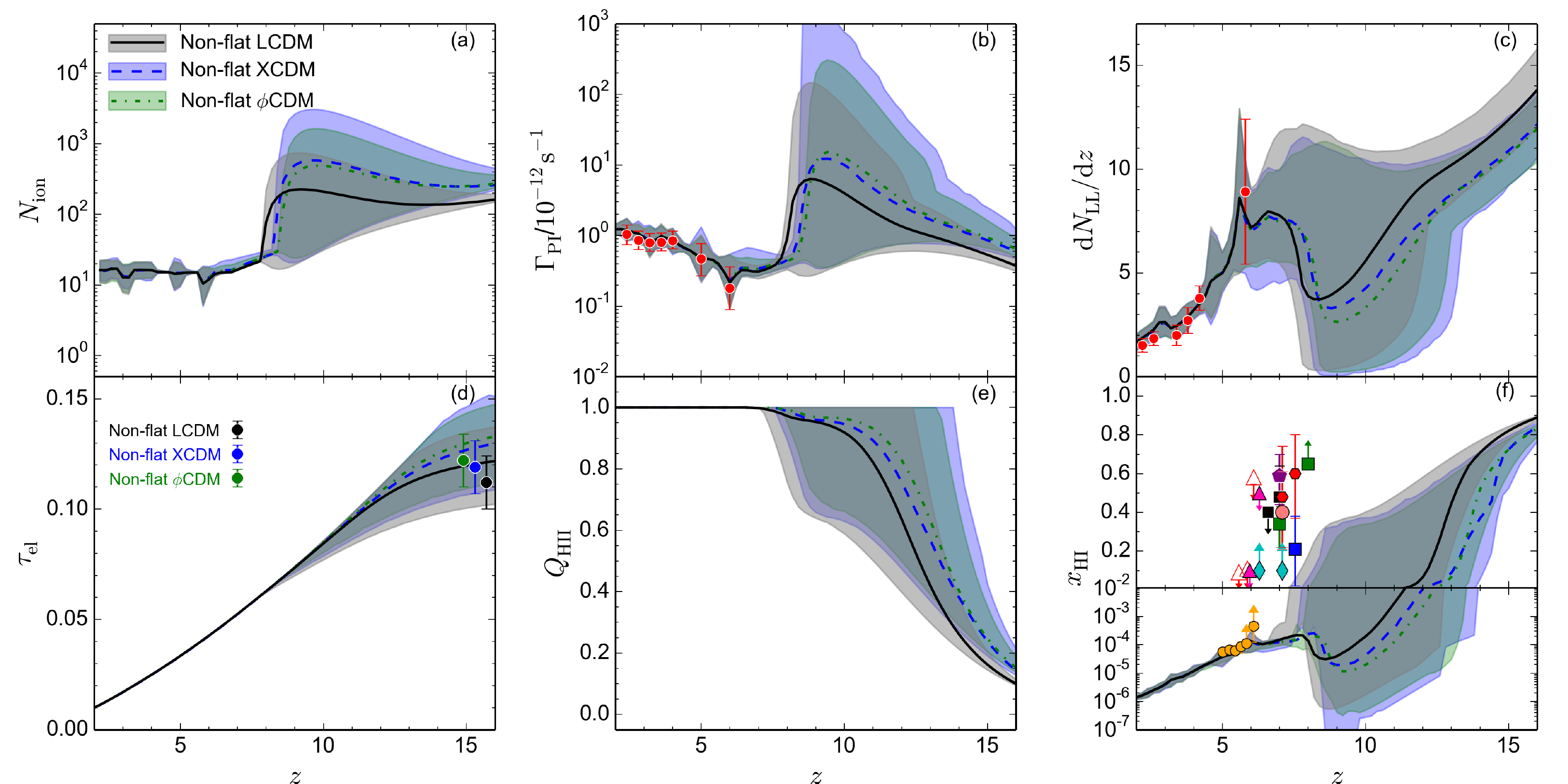}
  \caption{Same as Fig.~\ref{fig:MCMC-flat}, but now showing the constraints for the three non-flat inflation models.}
\label{fig:MCMC-nonflat}
\end{figure*}

We also perform a similar analysis for flat and non-flat XCDM and $\phi$CDM 
dynamical dark energy inflation models and these results are shown
in Figs.~\ref{fig:MCMC-flat} and \ref{fig:MCMC-nonflat} respectively. 
The evolution of all the quantities in these models have been indicated 
by dashed central lines with surrounding blue shaded 2-$\sigma$ regions 
(for XCDM) and dot-dashed central lines with surrounding green shaded 
2-$\sigma$ regions (for $\phi$CDM). For comparison, the $\Lambda$CDM models 
are shown by solid central lines with surrounding gray shaded 2-$\sigma$
regions.

The first thing to notice is that, irrespective of the nature of dark 
energy, models with similar reionization optical depths behave similarly. 
For all the flat models with relatively lower $\tau_{\rm el}$ 
($\approx0.06$---0.07), cosmic reionization can be accomplished by a 
single stellar population (Pop II stars). Although the mean evolution 
of $N_{\rm ion}(z)$ 
shows a slight increase at $z\gtrsim6$, a constant model is still permitted 
within its 2-$\sigma$ C.L. On the other hand, for all the non-flat models 
with $\tau_{\rm el}\approx0.11-0.12$, reionization has to be driven by other 
stellar populations, perhaps Pop III stars, at earlier epochs. The 
corresponding $\tau_{\rm el}$ data are shown in Panel (d) by the colored points
with error bars. As additional regions get ionized, the combined action of 
chemical and radiative feedback suppresses Pop III star formation and after 
that the Pop II stellar population dominates the reionization process at 
lower redshifts ($z\lesssim8$). Such models indicate a much faster evolution
of $Q_{\rm HII}(z)$ at initial stages, and then gradually approach towards the 
end-stage of the epoch of reionization. Unlike flat models, the non-flat ones 
hint at a much earlier and more extended reionization scenario 
that is completed around $z\approx7$. In fact, we find that higher the 
optical depth, the more extended is the reionization process. Also, the
neutral hydrogen fraction obtained from the non-flat models
is much smaller at higher redshifts ($z>7$) compared to the flat cases,
which makes these models likely to be
disfavored by most current high-redshift observations from distant QSOs,
GRBs and LAEs (see Panel (f) of Fig.~\ref{fig:MCMC-nonflat}).
However, the Planck 2018 reduction in $\tau_{\rm el}$ in the six parameter 
tilted flat $\Lambda$CDM inflation model by about 0.9-$\sigma$ will somewhat
help reconcile the non-flat model predictions with these observations.

\subsection{Evolution of the escape fraction}
\label{subsection:escapefrac}

\begin{figure*}
\centering
  \includegraphics[height=0.32\textwidth, angle=0]{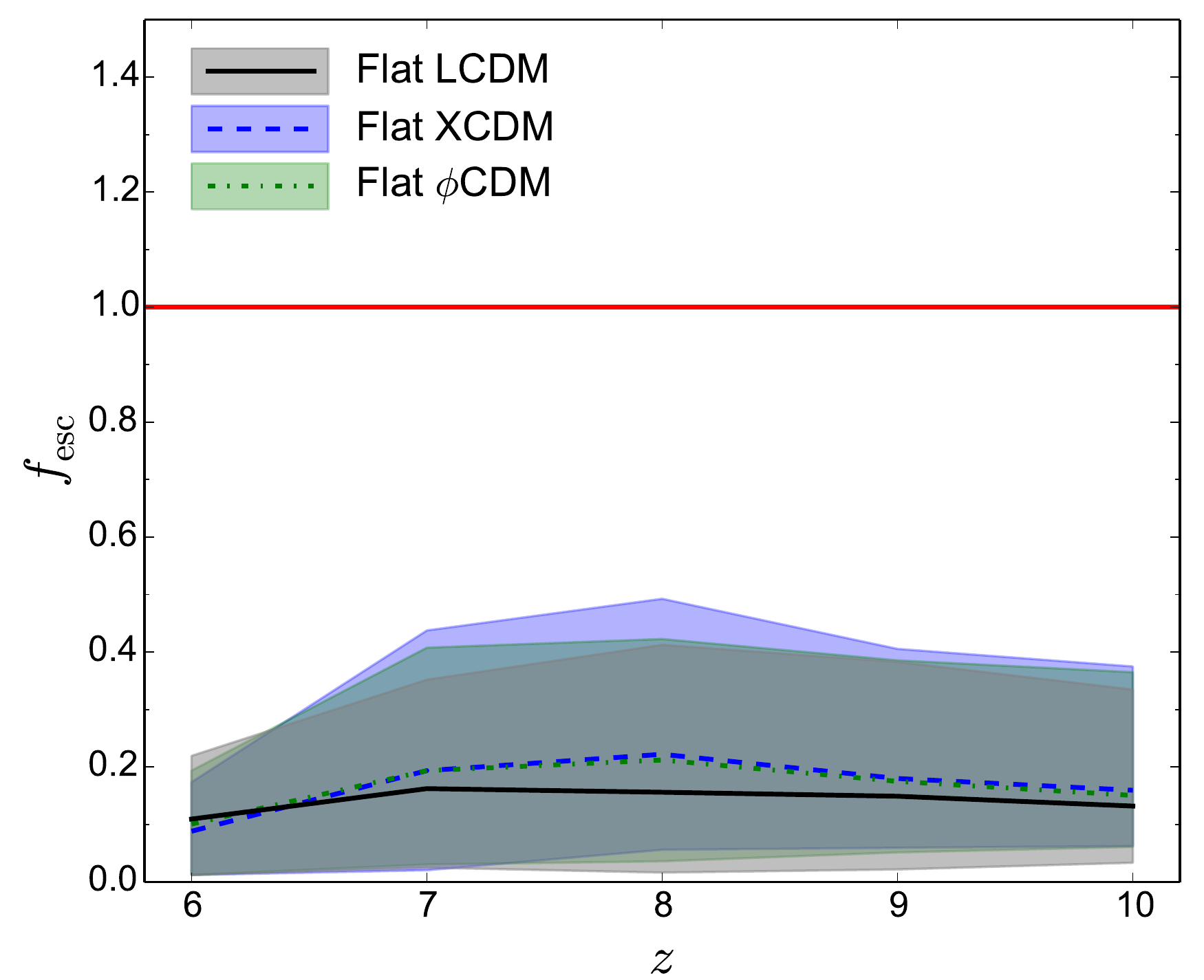}
  \hspace{10mm}
  \includegraphics[height=0.32\textwidth, angle=0]{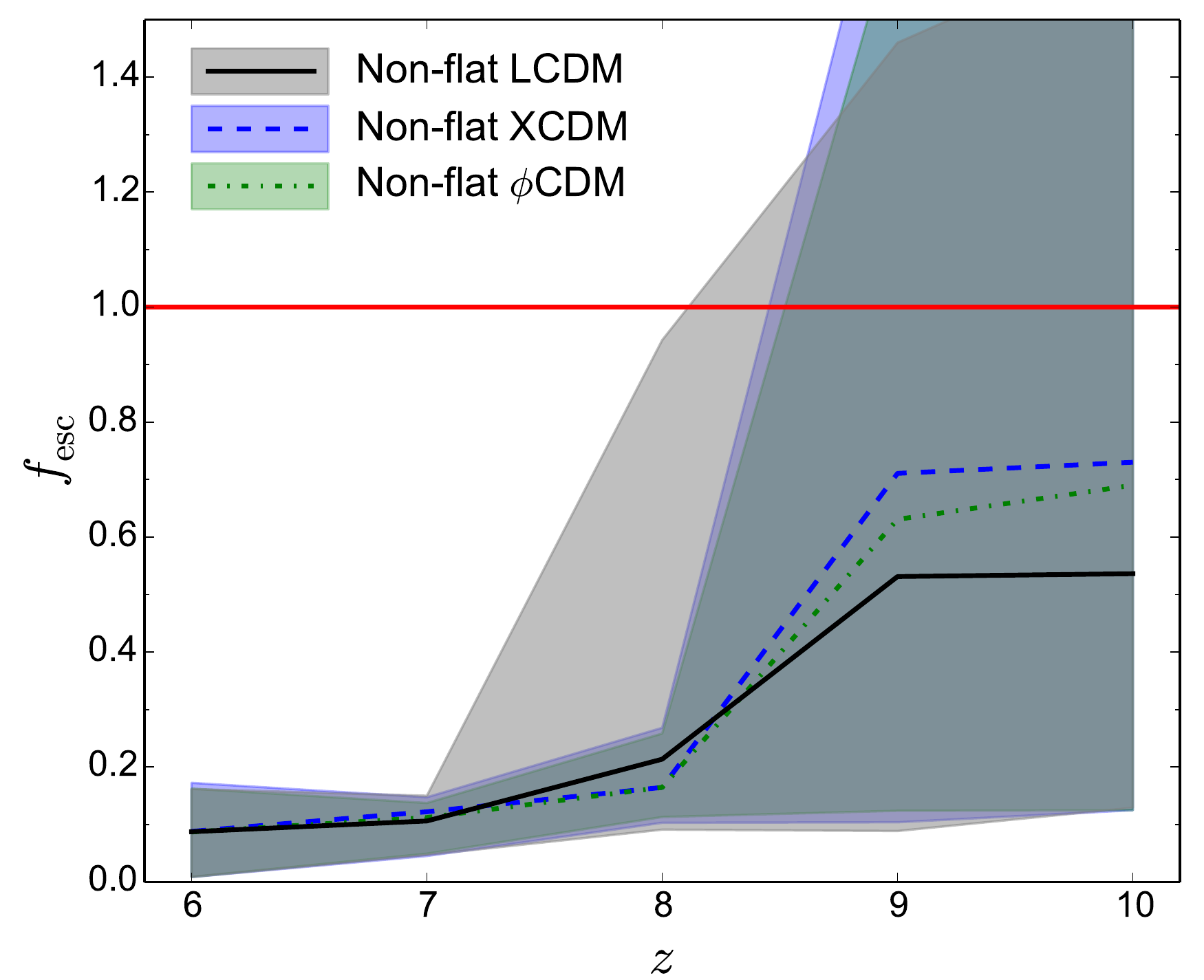}  
  \caption{Redshift evolution of the escape fraction $f_{\rm esc}(z)$ along with its
  2-$\sigma$ errors for all the flat ({\it left panel}) and non-flat ({\it right panel}) models
  considered in this work. The red solid lines in both panels correspond to an escape fraction of unity.}
\label{fig:escapefrac}
\end{figure*}

It is clear that all non-flat models, irrespective of the nature of dark energy,
predict very high $N_{\rm ion}$ values and their strong evolution with redshift.
This prompts us to check whether the escape fraction of ionizing photons from PopII
stars for such models becomes unrealistically high at some redshifts. This quantity
can be obtained by comparing the observed evolution of the high-$z$ galaxy UV Luminosity
Function (LF) with the predictions from our reionization models. There already exist
many in-depth studies for this, see e.g., \cite{2007MNRAS.377..285S,2009MNRAS.398.2061S,
2011MNRAS.412.2781K,mitra3,mitra4}. We refer the reader to those for the detailed methodology.

The basic idea is to compute the luminosity function $\Phi(M_{AB}, z)$, where $M_{AB}$ is
the absolute AB magnitude, from the luminosity at $1500$ $\mathring{\rm A}$ of a galaxy
which in turn depends on the star-forming efficiency $\epsilon_*$. We then vary $\epsilon_*$ 
as a free parameter to match the observed LFs at redshifts $z=6-10$.
Although not shown in the present paper, we find that a roughly constant
($\sim4\%$; best-fit value) $\epsilon_*$ is needed for all the flat and non-flat models
considered here (see Fig.~2 of Paper I; \citealt{mitra6}). We can then get limits for
$f_{\rm esc}$ at different redshifts from Eq.~(\ref{eq:Nion}) using the MCMC constraints
already obtained on the evolution of $N_{\rm ion}(z)$. Note that we take the value of
$N_{\gamma}$ as $\approx 3200$ which is appropriate for the PopII Salpeter IMF assumed here.

The redshift evolution of $f_{\rm esc}$ for different models are shown in Fig.~\ref{fig:escapefrac}.
The corresponding 2-$\sigma$ uncertainties, indicated by shaded regions in the plot, have been computed
using the quadrature method \citep{mitra3}. We find that the best-fit $f_{\rm esc}$ remains almost constant
at $\sim10-20\%$ for the whole redshift range in all the flat models ({\it left panel}). On the other
hand a strong redshift evolution of this quantity is required for the non-flat models ({\it right panel})
-- it can increase by a factor of $\sim 7$ from redshift $6$ to $10$. Moreover, if we consider their 2-$\sigma$
ranges, these non-flat models can lead to an impractically high $f_{\rm esc}$ ($\gtrsim1$) at
redshifts $z\gtrsim8$. The sole reason behind such trends is that the values of $N_{\rm ion}$ for non-flat models
can become as high as $\sim1000$ around $z\approx8-9$, considering their 2-$\sigma$ limits
(see panel (a) of Fig.~\ref{fig:MCMC-nonflat}), in order to produce very high optical depths
required for those models. On the other hand, the upper limits (2-$\sigma$) of escape fraction for the flat models having
relatively lower $\tau_{\rm el}$, can become at most $50\%$ for all the redshift ranges considered here.
Such one-to-one correspondence between strong redshift evolution of $f_{\rm esc}(z)$ and higher
reionization optical depths has also been reported earlier in many studies, see e.g.,
\cite{2012ApJ...746..125H,2012MNRAS.423..862K,mitra3}.

\section{Concluding remarks}
\label{sec:conclusions}

We have presented detailed statistical analysis of reionization 
in tilted flat and untilted non-flat $\Lambda$CDM, XCDM and $\phi$CDM 
inflation models. The cosmological parameters for these models are 
constrained by Planck 2015 TT + lowP + lensing CMB anisotropy and SNIa, BAO, 
$H(z)$, and growth rate data, using physically motivated inflation
power spectra for energy density inhomogeneities
\citep{2018arXiv180305522P,2018ApJ...868...83P}. For the non-flat models, these data prefer
mildly closed  models with $\Omega_{\rm k} \sim-0.006$ to $-0.008$. Although 
such models provide better fits to the observed low-$\ell$ temperature 
anisotropy $C_\ell$'s and weak-lensing $\sigma_8$ estimates, they provide
worse fits to the observed higher-$\ell$ temperature anisotropy $C_\ell$'s
and primordial deuterium abundances \citep{2018PASP..130k4001P}. These 
closed models also predict relatively higher reionization optical depth 
values ($\tau_{\rm el}\approx0.11$---0.12) compared to those obtained from 
the flat ones.
This could lead to a significantly different reionization scenario at higher redshifts $z>6$ in the non-flat cases.
To get constraints on reionization parameters, we decompose, an unknown but yet a very crucial quantity,
$N_{\rm ion}(z)$, the number of photons in the IGM per baryon, into its principal components and perform a thorough
MCMC analysis on the PCA modes using joint datasets of quasars and the corresponding $\tau_{\rm el}$ for each model.
Our analysis method is quite similar to that presented in Paper I \citep{mitra6}.

Our main findings, in summary, are:
\begin{itemize}
 \item We find that all six models behave very similarly in the lower 
redshift region ($z\lesssim6$) and can comfortably match all available 
observational data here, whereas at earlier epochs they differ significantly, 
as expected, due to the different optical depth values of the models.
 
 \item All the non-flat models, irrespective of their nature of dark energy, 
demand a strong redshift evolution in $N_{\rm ion}(z)$ at $z>6$ in order to 
produce the higher $\tau_{\rm el}$  values. This could hint at the possibility 
of reionization driven by early stellar sources like Pop III stars and/or 
a rapidly increasing star formation efficiency and/or photon escape fraction. 
On the other hand, a constant $N_{\rm ion}$ is  sufficient to explain 
reionization in flat models.
 
 \item Models with higher optical depths result in a relatively extended 
and early reionization completing around $7\lesssim z\lesssim13$ 
(from 2-$\sigma$ limits for $Q_{\rm HII}\sim1$). Also, such models predict a 
much lower neutral hydrogen fraction at higher redshifts ($z\gtrsim7$). Such 
small values, e.g. $<0.2$ at $z\sim8$, are clearly in tension with most of 
the current observational limits from distant Ly$\alpha$ emitters.
\end{itemize}

Another serious drawback of the non-flat models can be seen from the evolution 
of the photon escape fraction $f_{\rm esc}(z)$.
In Fig. \ref{fig:escapefrac}, we have shown that models with very large $N_{\rm ion}$
at higher redshifts give rise to an unrealistically high escape fraction at those redshifts.
We find that $f_{\rm esc}$ in the non-flat models can become $\gtrsim1$
even at $z\gtrsim8$, considering its 2-$\sigma$ limits. Such unphysical values of $f_{\rm esc}$
point to the possibility of ruling out those models.\footnote{As noted above, the most recent
\citet{2018arXiv180706209P} analyses result in $\tau_{\rm el}$ values around 1-$\sigma$ lower than
the Planck 2015 values we have used here. The lower $\tau_{\rm el}$ values will result in lower
$f_{\rm esc}$ values than what we have derived here.} 
On the other hand, a constant escape
fraction of $\sim10-20\%$ (best-fit) is adequate for all the flat models.
Note that, we did not include the high-redshift ($z>7$) observational bounds on $x_{\rm HI}$
in our current analysis as these are quite weak and largely model-dependent in nature
\citep{dayal1,2013MNRAS.429.1695B,2015MNRAS.452..261C,2016MNRAS.463.4019K,
2018MNRAS.479.2564W}. But even if we incorporate such data in the model to 
discard some of the very early reionization scenarios which are otherwise 
allowed in the non-flat cases, the evolution of various reionization 
quantities can become significantly non-monotonic,
which is quite unphysical in nature and thus should be disfavored.
For detailed discussion on this, we refer the reader to our Paper I \citep{mitra6}.

It is now well recognized that the high-redshift LAE data favor a late 
reionization \citep{2015MNRAS.446..566M, 2015MNRAS.452..261C} and the non-flat 
models having large $\tau_{\rm el}$  values clearly struggle to match these data.
This raises the possibility of ruling out such models in view of
those current observational limits.
Unrealistic escape fractions also push the non-flat models to the verge of 
being ruled out. We emphasize however that the lower Planck 2018 
$\tau_{\rm el}$ will alleviate some of this tension; it remains to be 
established whether this $\tau_{\rm el}$ reduction can alleviate all of 
the tension.

On the other hand, all the flat models, irrespective of the nature of dark 
energy, behave almost the same, making it difficult to distinguish between 
them by using available data. One has to rely on future observations of more
high-redshift quasars, LAEs, and possible detection of the redshifted 
21-cm signal from the epoch of reionization to resolve this issue, and 
possibly establish or rule out dark energy dynamics.

\section*{Acknowledgements}

This work was partially supported by DOE grant DE-SC0019038. 
C.-G.P.\ was supported by the Basic Science Research Program through the National Research Foundation of Korea (NRF)
funded by the Ministry of Education (No. 2017R1D1A1B03028384).




\bibliographystyle{mnras}
\bibliography{reionization-smitra} 

\bsp	
\label{lastpage}
\end{document}